\documentclass{aer} 

\usepackage[super,comma,round]{natbib}
\usepackage[figuresleft]{rotating}
\usepackage{txfonts}
\usepackage[hyphens]{url}
\usepackage{color}
\usepackage{eurosym}

\newcommand{\MARKI}[1]{#1}
\newcommand{\MARKII}[1]{#1}
\newcommand{\MARKIV}[1]{#1}

\jmonth{Month}
\jyear{YYYY}
\volume{XXX}
\issue{xxx}
\historydate{Received DD MM YYYY; revised DD MM YYYY; accepted DD MM YYYY}
\doi{10.1017/aer.2016.1} 

\begin{document}

\markboth{Janhunen et al.}{Steam balloon for rockets}


\title{Steam balloon concept for lifting rockets to launch altitude} 


\author{P.~Janhunen\email{pekka.janhunen@fmi.fi}
\\ P.~Toivanen\email{petri.toivanen@fmi.fi}
\\ K.~Ruosteenoja\email{kimmo.ruosteenoja@fmi.fi}
}

\affil{Finnish Meteorological Institute
\\ Helsinki\\ Finland}



\maketitle                   

\begin{abstract}
  Launching orbital and suborbital rockets from a high altitude is
  beneficial because of e.g.~nozzle optimisation and reduced drag.
  Aeroplanes and gas balloons have been used for the purpose.  Here we
  present a concept where a balloon is filled with pure water vapour
  on ground so that it rises to the launch altitude. The system
  resembles a gas balloon because no onboard energy source is carried
  and no hard objects fall down. We simulate the ascent behaviour of
  the balloon. In the baseline simulation, we consider a 10 tonne
  rocket lifted to an altitude of 18 km. We model the trajectory of
  the balloon by taking into account steam adiabatic cooling, surface
  cooling, water condensation and balloon aerodynamic drag. The
  required steam mass proves to be only 1.4 times the mass of the
  rocket stage and the ascent time is around 10 minutes.  For small
  payloads, surface cooling increases the relative amount of steam
  needed, unless insulation is applied to the balloon skin.  The
  ground-filled steam balloon seems to be an attractive and
  sustainable method of lifting payloads such as rockets into high
  altitude.
\end{abstract}


\section*{NOMENCLATURE} 
\begin{deflist}
\listterm{$A$}{Cross-sectional area of balloon (m$^2$)}
\listterm{$A_\mathrm{skin}$}{Atmosphere facing area of steam-filled volume (m$^2$)}
\listterm{$C_D$}{Drag coefficient}
\listterm{$c_p$}{Specific heat capacity of air in constant pressure (J kg$^{-1}$ K$^{-1}$)}
\listterm{$F_D$}{Drag force (N)}
\listterm{$f_r$}{Rejected fraction of condensed water}
\listterm{$g$}{Acceleration due to gravity, $g=9.81$ m s$^{-2}$}
\listterm{$k$}{Thermal conductivity of air (W K$^{-1}$ m$^{-1}$)}
\listterm{$k_B$}{Boltzmann constant (J K$^{-1}$)}
\listterm{$L$}{Length scale (m)}
\listterm{$L_z$}{Height of cylinder approximating balloon (m)}
\listterm{$m_\mathrm{pay}$}{Payload mass (kg)}
\listterm{$m_\mathrm{s}$}{Steam mass (kg)}
\listterm{$n$}{Atmospheric number density (m$^{-3}$)}
\listterm{$\mathrm{Nu}_L$}{Nusselt number corresponding to length scale $L$}
\listterm{$P$}{Pressure (Pa)}
\listterm{$\mathrm{Pr}$}{Prandtl number of air, $\mathrm{Pr}=\nu/\alpha$}
\listterm{$r$}{Radius of balloon (m)}
\listterm{$\mathrm{Re}_L$}{Reynolds number or air corresponding to length scale $L$}
\listterm{$T$}{Temperature (K)}
\listterm{$u_\mathrm{conv}$}{Convective cooling rate of balloon skin (W m$^{-2}$)}
\listterm{$u_\mathrm{rad}$}{Radiative cooling rate of balloon skin (W m$^{-2}$)}
\listterm{$\varv$}{Vertical velocity (m s$^{-1}$)}
\listterm{$V$}{Steam volume (m$^3$)}
\end{deflist}


\subsection*{Greek symbols}
\begin{deflist}
\listterm{$\alpha$}{Thermal diffusivity of air (m$^2$ s$^{-1}$)}
\listterm{$\gamma$}{Adiabatic index of water vapour, $\gamma=1.324$}
\listterm{$\epsilon$}{Thermal infrared emissivity of balloon skin}
\listterm{\MARKIV{$\eta$}}{\MARKIV{Relative lift}}
\listterm{$\nu$}{Kinematic viscosity of air (m$^2$ s$^{-1}$)}
\listterm{$\mu$}{Dynamic viscosity of air, $\mu=\rho\nu$ (Pa s)}
\listterm{$\rho$}{Atmospheric mass density (kg m$^{-3}$)}
\listterm{$\sigma_\mathrm{SB}$}{Stefan-Boltzmann constant, $\sigma_\mathrm{SB}=5.67\cdot 10^{-8}$ W m$^{-2}$ K$^{-4}$}
\end{deflist}

\section{INTRODUCTION}

Rockets benefit from launching at a high
altitude\cite{SarigulKlijnEtAl2005}. High launch altitude reduces air
drag and improves possibilities to optimise the nozzle. Launch above
ground may also be desirable for other reasons such as avoidance of
rocket noise and safety risks on ground.

Aeroplanes have been used for air launchers such as
Pegasus\cite{Pegasus1994}, Stratolaunch and Launcher
One\cite{NiederstrasserAndFrick2015}. Using a tailor-made plane such
as White Knight 2 developed by Scaled Composites company is a
possible, but rather expensive option. One could also use an existing
military transporter such as Boeing C-17 Globemaster
III\cite{SarigulKlijnEtAl2005} or a civilian plane modified to carry
or tow the rocket. However, the loaded maximum altitude of such
aeroplanes is typically $\sim 10-13$ km, while nozzle vacuum
performance would benefit from somewhat larger altitude such as
$\sim 18$ km. Also, if legislation or other considerations make it
impractical to use the modified plane for other purposes, the
per-launch capital and maintenance costs can become significant,
unless the launch frequency is high.

Helium balloons can reach higher altitude than standard aeroplanes,
and they have been used for launching sounding rockets in the
past\cite{VanAllen1994} and are planned to be used again, e.g., by
Zero2infinity company's Bloostar launcher\cite{Bloostar2017}. However,
helium is a moderately expensive gas. It is also a non-renewable
resource that is currently obtained as a byproduct of natural gas
extraction\cite{Mullins1960}. In the future, the price of helium may
increase, the price can fluctuate, and availability issues are
possible. Thus there is a motivation to look for possible alternative
lifting gases.

Hydrogen is the other well-known lifting gas (Table \ref{tab:gases}).  Although readily
available, hydrogen is likewise not very cheap. Its energy cost is
relatively high: to levitate a 1 kg mass requires an amount of
hydrogen which would release 10.5 MJ of energy upon combustion, and
typically even more energy is needed to produce it. Hydrogen
is also somewhat complicated to transport and handle. During the balloon
filling operation, hydrogen carries a risk of fire if, e.g., the balloon
skin is punctured so that a hydrogen stream escapes into the
atmosphere and produces a flammable mixture that can be ignited,
e.g., by a static electricity spark. Although the fire risk can be
mitigated by quality assurance practices, running such practices
incurs additional financial cost.

\begin{table}[h]
\tabcolsep7.5pt
\caption{Properties of selected lifting media.}
\label{tab:gases}
\centering
\begin{tabular}{llllll}
\textbf{Gas} & \textbf{Molecular}  & \textbf{Relative} & \textbf{Flam-} & \MARKII{\textbf{Price\MARKII{$^{a}$} per}} & \textbf{Long term} \\
& \textbf{weight} & \textbf{lift} & \textbf{mable} & \MARKII{\textbf{levitated tonne}} & \textbf{availability} \\[6pt]
Vacuum   & -  & 100\,\% & - & - & - \\
Hydrogen & 2  & 93\,\% & Yes & \MARKII{900\,\euro} & Good\\
Helium   & 4  & 86\,\% & No  & \MARKIV{3000}\,\euro & Limited \\
Saturated steam & 18 & 51-59\,\%$^{b}$ & No & \MARKII{70\,\euro$^{c}$} & Good \\
Methane  & 16 & 45\,\% & Yes & \MARKII{500\,\euro} & Good \\
Neon     & 20 & 31\,\% & No  & \MARKII{90000\,\euro} & Good \\
Hot air  & 29 & 22\,\%$^{d}$ & No & \MARKII{5\,\euro$^{c}$} & Good \\
\end{tabular}
\begin{tabnote}
\MARKII{$^{a}$Exemplary prices from Internet shops and analyst reports,
gathered in August 2018.}\\
$^{b}$The range is due to the dependence of temperature and pressure on altitude.\\
\MARKII{$^{c}$Energy cost assuming methane as fuel, multiplied by 2 to approximate inefficiencies.}\\
$^{d}$For +100 $^{\circ}$C air at +20 $^{\circ}$C ambient.\\
\end{tabnote}
\end{table}

Natural gas or methane could be used as a lifting gas as well, but
like hydrogen this gas is flammable. Methane is a greenhouse gas whose
escape into the atmosphere should be limited. Methane is also rather
energy intensive in the sense that levitating a 1 kg mass requires an
amount of methane that would produce as much as 68.3 MJ of heat if
burned in a power plant instead of being used as a lifting gas. In
this regard, methane is 6.5 times more energy intensive than
hydrogen. The main reason is methane's much larger molecular weight
which implies that a larger mass of the gas is needed to produce the
same lift (Table \ref{tab:gases}).

Water vapour (steam) is a potent lifting gas (Table
\ref{tab:gases}). Its lifting capacity is a result of the lower
molecular weight (18 versus 29 in the atmosphere) and the higher
temperature compared to the ambient air. At sea level at room
temperature, +100 $^{\circ}$C steam has 51\,\% of the lifting capacity
of vacuum while unheated helium has 86\,\%. Steam was proposed as a
lifting gas already 200 years ago \cite{Cayley1816} and has been
suggested thereafter several times for
balloons\cite{Alcock1961,Giraud1993} and
airships\cite{Erdmann1909,Papst1969}. More recently, a low altitude
balloon ``HeiDAS'' that used both hot air and steam was constructed
\cite{BormannEtAl2003}.  The HeiDAS balloon resembles a hot-air
balloon as it carries a boiler to re-evaporate the condensed
water. Using steam enables making the balloon's volume smaller by
factor $\sim 2.5$ than a hot-air balloon. On the other hand, the
needed water collector and boiler is a more complex apparatus than the
simple propane burner typically used in hot-air balloons.

In this paper we consider the possibility to use a steam balloon that
does not need any onboard energy source or other onboard device. The
balloon is filled by steam on ground. It then rises together with the
payload to the target altitude. The rise time is relatively short,
typically of the order of 10 minutes as will be seen below, so that
the steam does not have time to cool significantly by heat loss
through the balloon envelope. The device resembles a high-altitude
zero pressure gas balloon\cite{RainwaterAndSmith2004}, except that it is filled with
steam instead of helium.  The underlying physics differs from the
previously considered low altitude steam balloons
\cite{Cayley1816,BormannEtAl2003}, because in addition to surface
cooling, volumetric adiabatic cooling of the steam also becomes
important. To the best of the authors' knowledge, the possibility of
filling a balloon with steam on ground and letting it rise to high
altitude has not been treated in literature before.

As a baseline, we consider the task of lifting a 10 tonne payload into 18
km altitude. We also explore variations of these
parameters. After the payload is released, the baseline idea is to
release the steam from the top of the balloon so that the balloon
falls down, typically into the sea from which it is recovered by a boat
for material recycling or for re-flight.

The structure of the paper is as follows. We first present a
mathematical model for the steam balloon and make vertical trajectory
simulations to find out the required steam mass. Then we explore the
dependence of the steam mass on the various parametres and alternative
assumptions. Next, we discuss three design variants and look into some
insulation options. The paper ends with a discussion and conclusions
section.

\section{STEAM BALLOON MODEL}

\begin{figure}[htpb]
\centering
\begin{minipage}[t]{0.45\textwidth}
\centering
\includegraphics[height=2.5in]{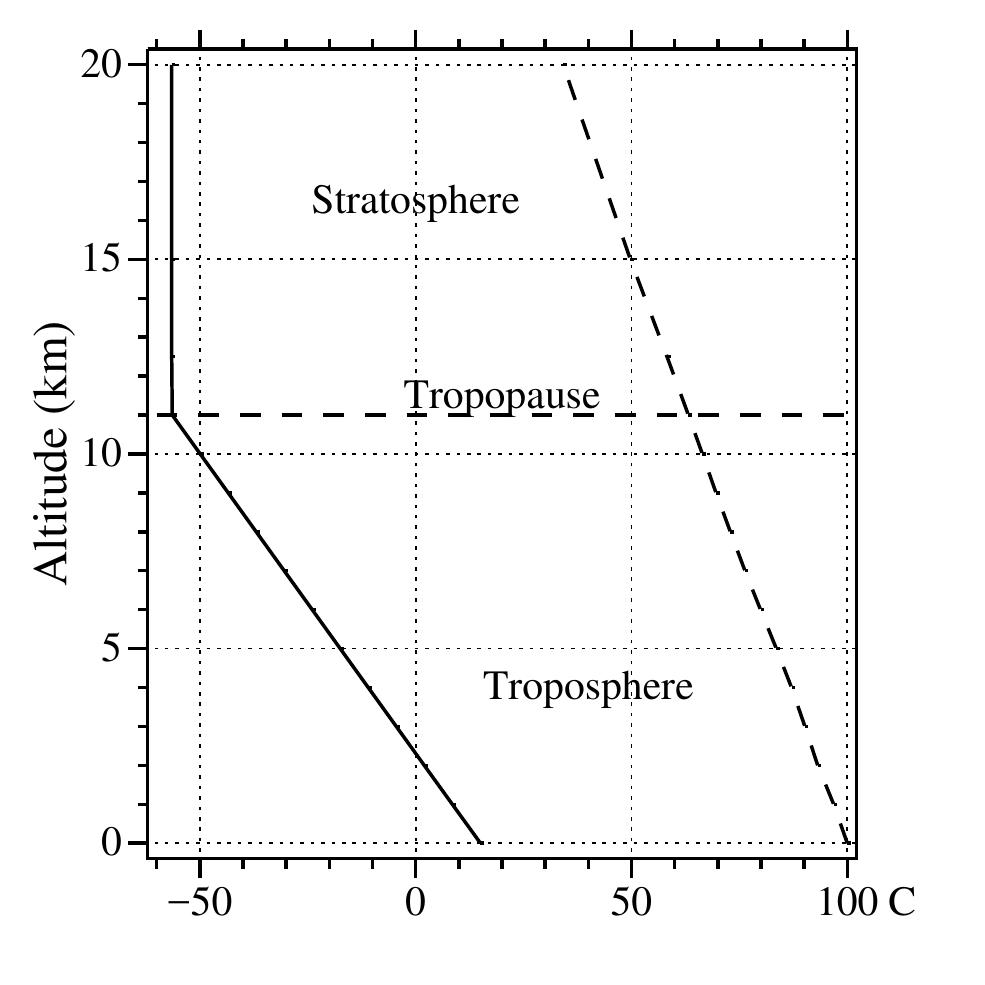}
\caption{Model atmosphere temperature (solid) and steam
  dewpoint temperature (dashed).}
\label{fig:T}
\end{minipage}\hfill
\begin{minipage}[t]{0.45\textwidth}
\centering
\includegraphics[height=2.5in]{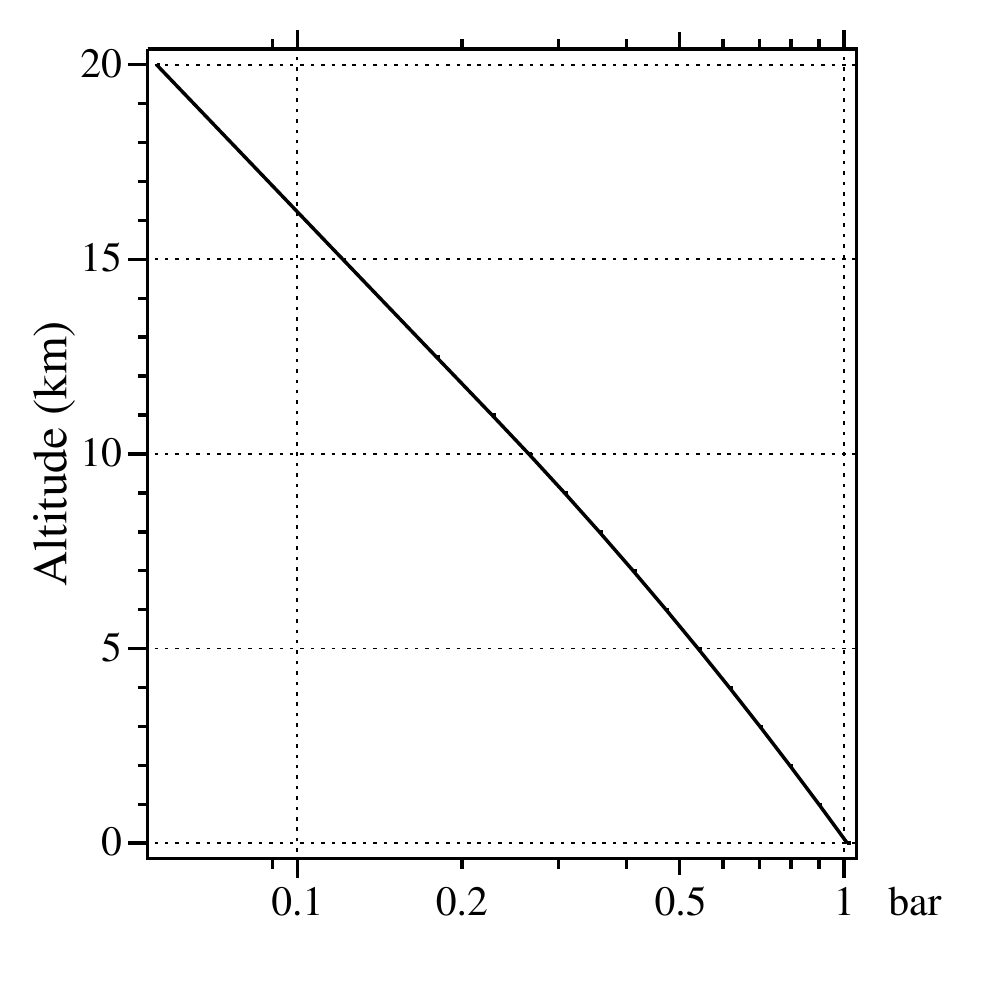}
\caption{Model atmosphere pressure profile.}
\label{fig:P}
\end{minipage}
\end{figure}

For the vertical temperature profile, we use a simple atmospheric
model\cite{USStandardAtmosphere1976} in which the temperature drops
linearly from +20 $^{\circ}$C until the tropopause at 11 km and is
then constant -55 $^{\circ}$C in the stratosphere. Dependence of the
performance on latitude and the local climate and weather is beyond
the scope of this paper. Steam (i.e., 100\,\% water vapour) has a
dewpoint temperature that depends on the ambient pressure. The atmospheric
temperature and the steam dewpoint temperature are shown in
Fig.~\ref{fig:T} while the ambient atmospheric pressure is given in
Fig.~\ref{fig:P}.

When the steam balloon rises, the steam expands and cools
adiabatically with an adiabatic index of $\gamma=1.324$, following the
adiabatic equation of state $P \sim n^\gamma$, or $n \sim
P^{1/\gamma}$. Then, using the ideal gas law $P=n k_B T$, the temperature is
proportional to
\begin{equation}
T \sim \frac{P}{n} \sim \frac{P}{P^{1/\gamma}} = P^{1-\left(1/\gamma\right)}.
\label{eq:adiabatic}
\end{equation}
However, when the steam cools below the dewpoint (Fig.~\ref{fig:T} dashed line)
by Eq.~(\ref{eq:adiabatic}), condensation occurs until the released
latent heat is sufficient to keep the temperature at the dewpoint. The
steam dewpoint drops more slowly than the ambient atmospheric
temperature (Fig.~\ref{fig:T}). For example, at the 18 km altitude
(pressure 75.65 hPa) the dewpoint is still +41 $^{\circ}$C. Only at 32
km altitude it drops below the freezing point.

The steam is cooled not only by adiabatic expansion, but also by
convective and radiative heat transfer through the balloon skin. Water
condensed on the inner surface keeps, by the release of
latent heat, the surface near the internal steam temperature.

We simulate the vertical trajectory of the balloon and its payload by
modelling the steam thermodynamics, surface cooling due to radiation and
convection and the aerodynamic drag that tends to slow down the
balloon ascent. To minimise the drag, the balloon has a streamlined
droplet-like shape. The faster the balloon rises, the less its
performance is affected by surface cooling. As the balloon rises,
steam condenses into water both volumetrically as cloud droplets and
on the inner surface. In the baseline case we assume, conservatively,
that none of the condensed water is rejected from the system. In
reality, at least part of the water would rain and trickle down and
would exit from the bottom opening.  The water does not freeze because
the dewpoint of pure steam is continuously much above the freezing
point. \MARKI{Moreover, risks of mechanical integrity due to ice formation are unlikely because
  even if the condensed water would freeze completely, the resulting ice layer
  would be only 0.25 mm thick.}

The balloon is pushed upwards by the lift of the steam. It is pulled
downwards by the gravitational weight contributed by the steam mass,
the non-rejected condensed water mass, the mass of the balloon itself
and the payload. The ascending motion is resisted by the aerodynamic drag,
\begin{equation}
F_D = \frac{1}{2} C_D \rho \varv^2 A
\end{equation}
where $A=\pi r^2$ is the balloon's cross-sectional area and $C_D$ is
the drag coefficient. The balloon's radius $r$ is chosen so that the
initial steam volume is equal to the volume of the \MARKII{sphere}
$(4/3)\pi r^3$. When the balloon rises, the steam expands, but we keep
a constant $r$ because we assume that the expansion goes into
lengthening the balloon, i.e., into making it a more elongated
droplet. This can be realised, e.g., if the balloon has an elongated
shape, but is initially filled only partially so that the lower part
hangs like a curtain (i.e., a similar strategy to
high-altitude zero pressure helium balloons\cite{RainwaterAndSmith2004}). We discuss different design options in section \ref{sect:variants} below.

The drag coefficient of a streamlined, droplet-like shape (at high Reynolds number
$\sim 10^7$ which is relevant in our case) is about 0.04
\cite{Hoerner1965}. The optimal shape has length about three
times the diameter\cite{Hoerner1965} so it is not far from typical balloon shape. The
optimum is broad, i.e., the drag coefficient is not very sensitive to the
exact length to diameter ratio. As a baseline we assume a drag coefficient of $C_D=0.08$, i.e.~twice
the theoretical optimum. We do this to account
qualitatively for any suboptimality in balloon shape \MARKI{plus the
  drag of the} the \MARKI{carried} rocket
and the ropes from which the \MARKI{rocket} hangs. \MARKI{Owing to
  their small size compared to the balloon, the
  contribution of the rocket and the ropes to the drag is rather
  negligible, of the order of a few percent typically.} The value $C_D=0.08$
is only the baseline alternative, and below we simulate a range of potential drag coefficient values.

Skin cooling is estimated by summing the radiative and convective contributions.
The radiative cooling rate (power per unit area, W m$^{-2}$)
is given by
\begin{equation}
u_\mathrm{rad} = \epsilon \sigma_\mathrm{SB} \left(T^4-0.5\,T_\mathrm{Earth}^4\right)
\label{eq:radiative}
\end{equation}
where $\epsilon$ is the thermal infrared emissivity and
$T_\mathrm{Earth}$ is Earth's effective infrared temperature for which
we use the value 255 K. The factor $0.5$ in front of
$T_\mathrm{Earth}$ comes from the fact that the topside of the balloon
radiates to dark space after reaching higher altitudes. This estimate
somewhat overestimates the radiative cooling rate, i.e., it is
a conservative assumption regarding the lifting performance, because at low
altitudes the effective Earth radiative temperature is higher and
infrared radiation is received also from above. \MARKII{Overall in
  this paper, in places where accurate modelling is hard, we prefer
  approximations that stay on the conservative side, i.e., that do not
  overestimate the lifting capability of the balloon.}

The lifting gas (steam) first tends to fill the top part of the balloon. When
the balloon goes up, the steam expands due to the decreasing pressure,
progressively filling a greater portion of the balloon volume.
If the balloon is already full, steam starts to leak out from the bottom.
We approximate the atmosphere facing surface area of the
steam volume by a vertical cylinder whose radius is $r$ (radius of a
sphere whose volume equals the initial steam volume) and whose
\MARKII{volume $V$ equals the expanded volume of the steam. Then the
  cylinder} height (length) is $L_z=V/A$ where 
$A=\pi r^2$ \MARKII{is the cross-sectional area of the sphere}. Then the
cylinder area is \MARKII{given by}
\begin{equation}
A_\mathrm{skin} = 2\pi r^2 + 2 \pi r L_z.
\end{equation}
The cylinder approximation is used for estimating the skin area
in a simple way. The total radiative cooling power is equal to
$u_\mathrm{rad} A_\mathrm{skin}$.

\begin{figure}[htpb]
\centering
\includegraphics[width=3in]{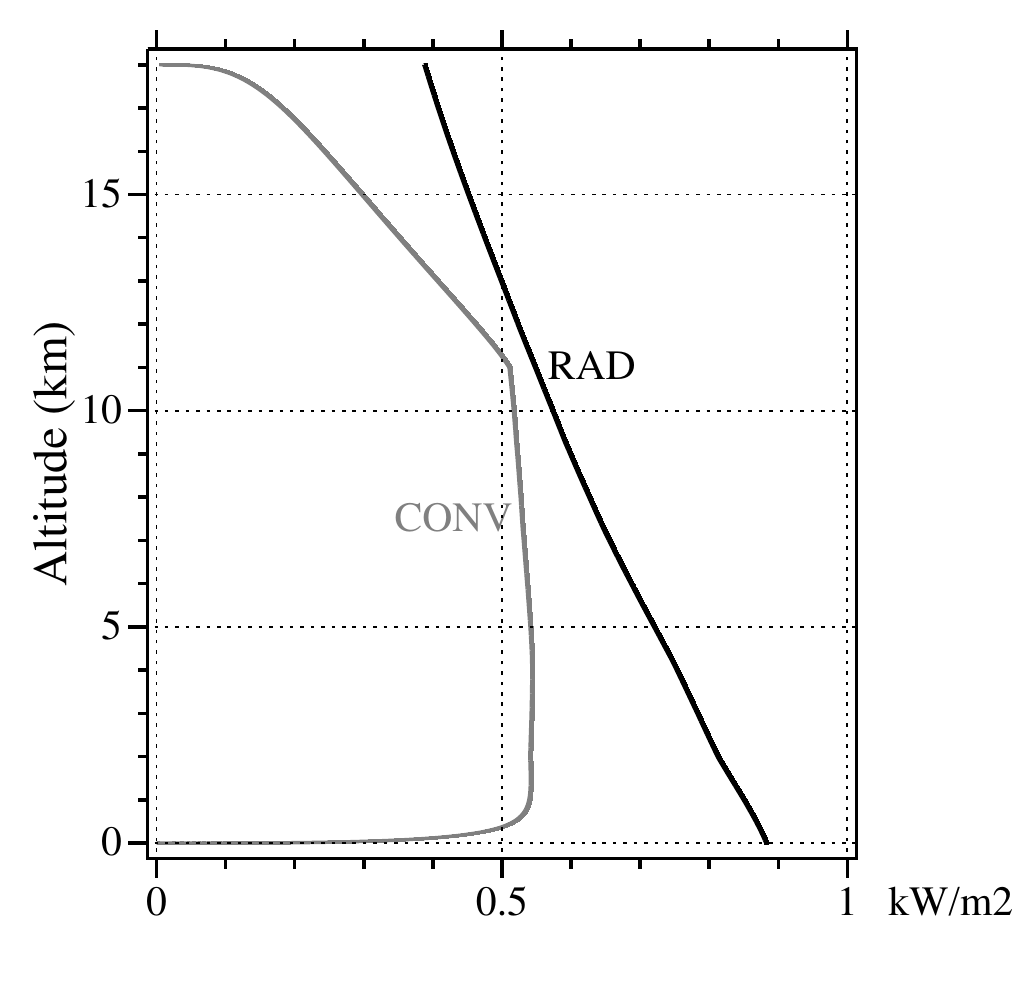}
\caption{The radiative \MARKII{(black)} and convective \MARKII{(grey)} cooling rates of the balloon skin
  in the baseline case.}
\label{fig:Heatflux}
\end{figure}

The convective cooling rate is more difficult to estimate. We use an
empirical approximation \cite{Whitaker1972} which is detailed in the
Appendix. The radiative and convective cooling rates as a function of
altitude are shown in Fig.~\ref{fig:Heatflux}. According to our
approximate calculations, the convective rate is smaller than the
radiative one at all altitudes.

Table \ref{tab:params} lists the main parameters of the balloon device
in the baseline simulation. 

\begin{table}[h]
\tabcolsep7.5pt
\caption{Device parameters in baseline case.}
\label{tab:params}
\centering
\begin{tabular}{lll}
\textbf{Parameter} & \textbf{Symbol} & \textbf{Value} \\[6pt]
Payload & $m_\mathrm{pay}$ & $10^4$ kg \\
Target altitude & & 18 km \\
Drag coefficient & $C_D$ & 0.08 \\
Rejected fraction of condensed water & $f_r$ & 0 \\
Thermal emissivity of the balloon skin & $\epsilon$ & 0.9 \\
Initial steam temperature & & +100$^{\circ}$C \\
Internal-external pressure difference & & 3 hPa \\
Balloon mass & & 0.1 times the initial steam mass \\
Initial steam mass & $m_\mathbf{s}$ & $14.2 \cdot 10^3$ kg (from the simulation) \\
\end{tabular}
\end{table}

\section{SIMULATION RESULTS}

\begin{figure}[htpb]
\centering
\begin{minipage}[t]{0.45\textwidth}
\centering
\includegraphics[height=2.7in]{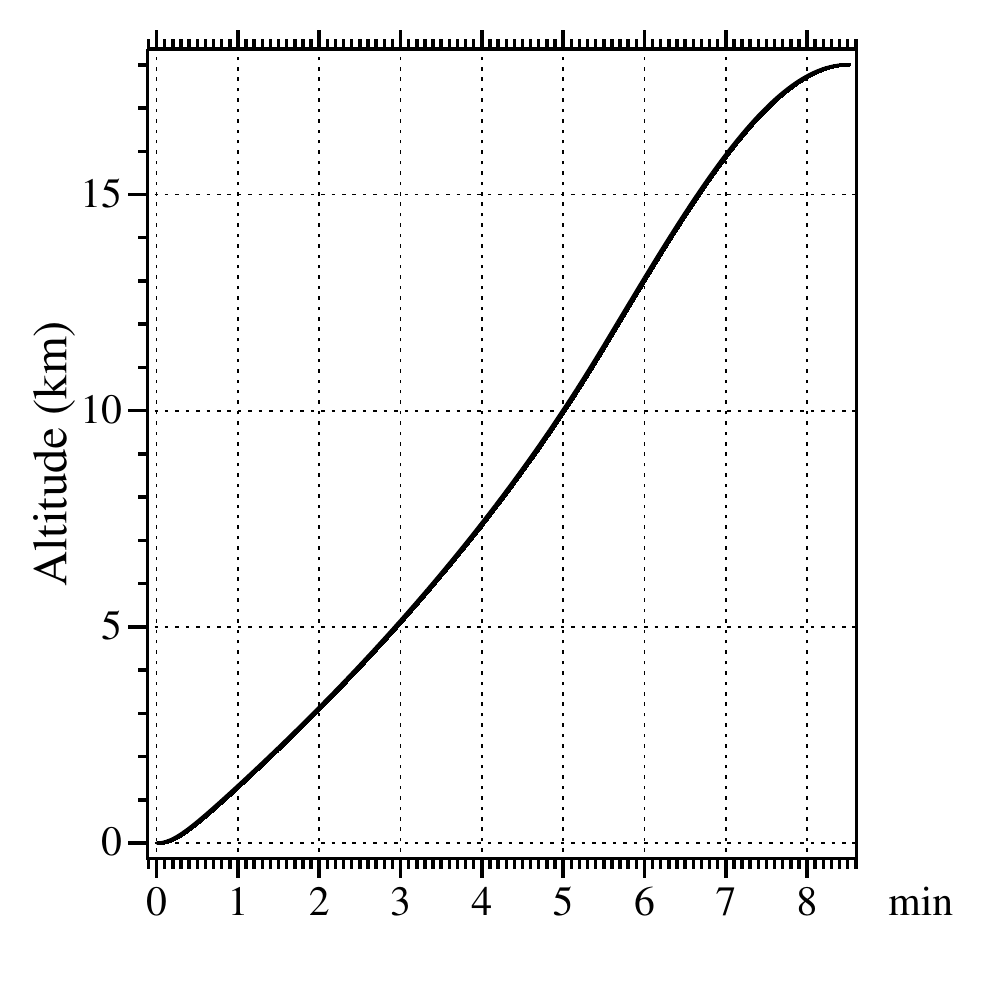}
\caption{Altitude as function of time\goodbreak in the baseline simulation.}
\label{fig:Alt}
\end{minipage}\hfill
\begin{minipage}[t]{0.45\textwidth}
\centering
\includegraphics[height=2.7in]{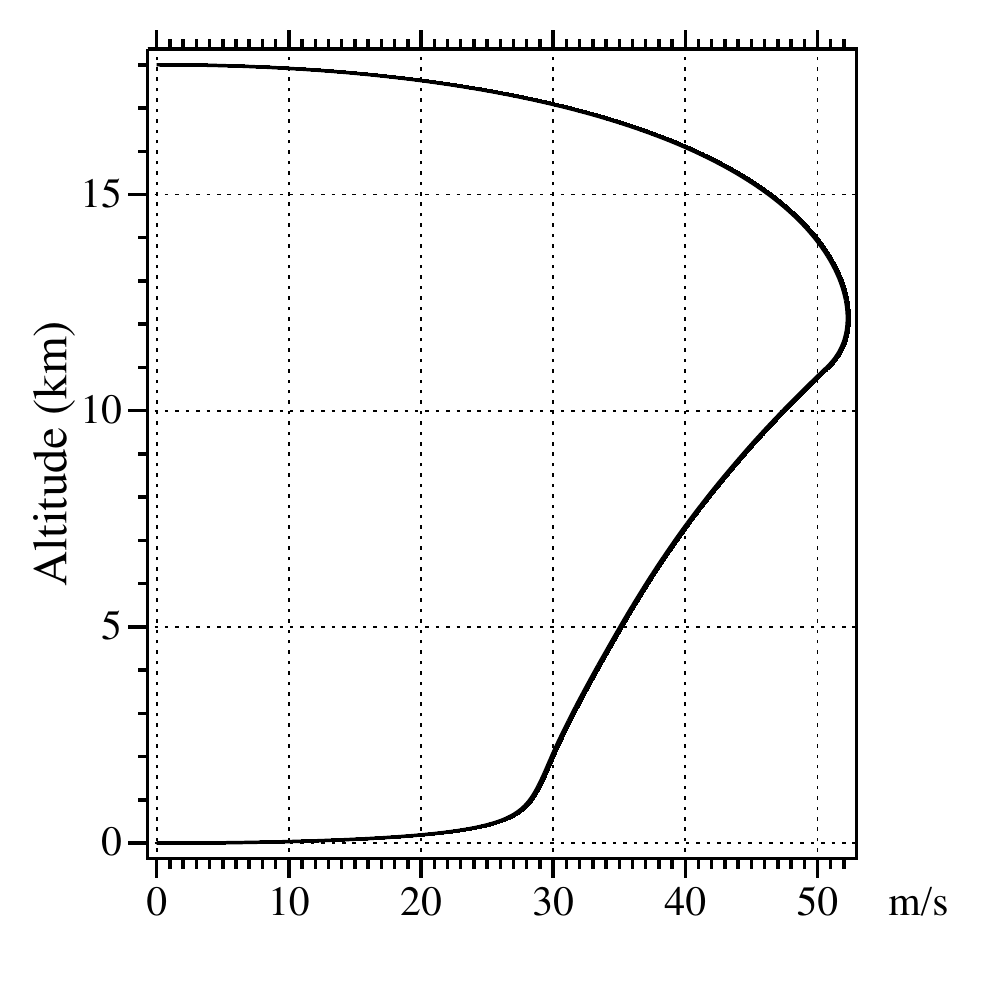}
\caption{Upward speed of the balloon\goodbreak in the baseline simulation.}
\label{fig:Speed}
\end{minipage}
\end{figure}

The trajectory and the upward speed of the balloon in the baseline
simulation (Table \ref{tab:params}) is shown in Figs.~\ref{fig:Alt}
and \ref{fig:Speed}. The balloon
\MARKI{takes off with a maximum acceleration of 0.15\,$g$}, and at $\sim 1$ km altitude the ascending speed
becomes drag-limited. The speed continues to increase as the air becomes
thinner and the internal minus external temperature difference
increases. After passing the tropopause, the speeds starts to decrease
because the ambient temperature no longer decreases while the internal
temperature continues to do so. The balloon reaches the maximum
altitude, which is by design 18 km in this case, and would then start
to descend. In the baseline case, the lifting from sea level to 18 km
altitude takes 8.5 min. At the maximum altitude, the payload rocket is
dropped and launched.

\MARKIV{

The simulated ascent speed of the balloon (up to 50 m/s,
Fig.~\ref{fig:Speed}) is much higher than, e.g., for weather
balloons. In what follows, we show that this is mainly a consequence of the large
size of the balloon. Assuming a spherical shape for simplicity, in steady state the
net lift is given by\cite{GalliceEtAl2011}
\begin{equation}
F_\mathrm{lift} = \eta \cdot \frac{4}{3} \pi r^3 \rho g
\end{equation}
where $\rho$ is the density of the ambient air and $\eta$ is the relative lift parameter,
which is a number between 0 and 1, 1 corresponding to an ideal massless balloon filled with weightless
lifting medium and 0 to zero net lift. The drag is given by\cite{GalliceEtAl2011}
\begin{equation}
F_\mathrm{drag} = C_D \cdot \frac{1}{2} \rho v^2 \pi r^2.
\end{equation}
Equating $F_\mathrm{lift}$ and $F_\mathrm{drag}$ and solving for the
equilibrium ascent speed $v$ yields\cite{GalliceEtAl2011}
\begin{equation}
v = \sqrt{\frac{8}{3} \frac{\eta r g}{C_D}}.
\label{eq:v}
\end{equation}
Thus the ascent speed is proportional to $\sqrt{r/C_D}$. In our case,
$r$ is much larger than the typical radius of weather balloons. Also, because of the
large size, the Reynolds number is large, above the drag crises value,
which keeps $C_D$ small. Additionally, steam cools slower than
adiabatically because of the latent heat released by
condensation. This effect is peculiar to steam balloons, and it boosts
the ascent speed especially during the upper troposphere portion of the climb.


Although the ascent speed is rather high, the overall drag force is
necessarily lower than the lift because most of the lift goes to
suspending the payload. Hence, if the balloon withstands its own lift
when anchored on ground, it also survives the aerodynamic
drag force during the ascent. Wake turbulence is a potential concern
because it might cause fluttering of the fabric. However, e.g.~in parachutes used for
air-dropping heavy payloads and for space capsule supersonic re-entries, such
issues have been successfully overcome. In any case, quantitative analysis of
structural engineering of the balloon is beyond the scope of the
present work.



}

\begin{figure}[htpb]
\centering
\includegraphics[width=3in]{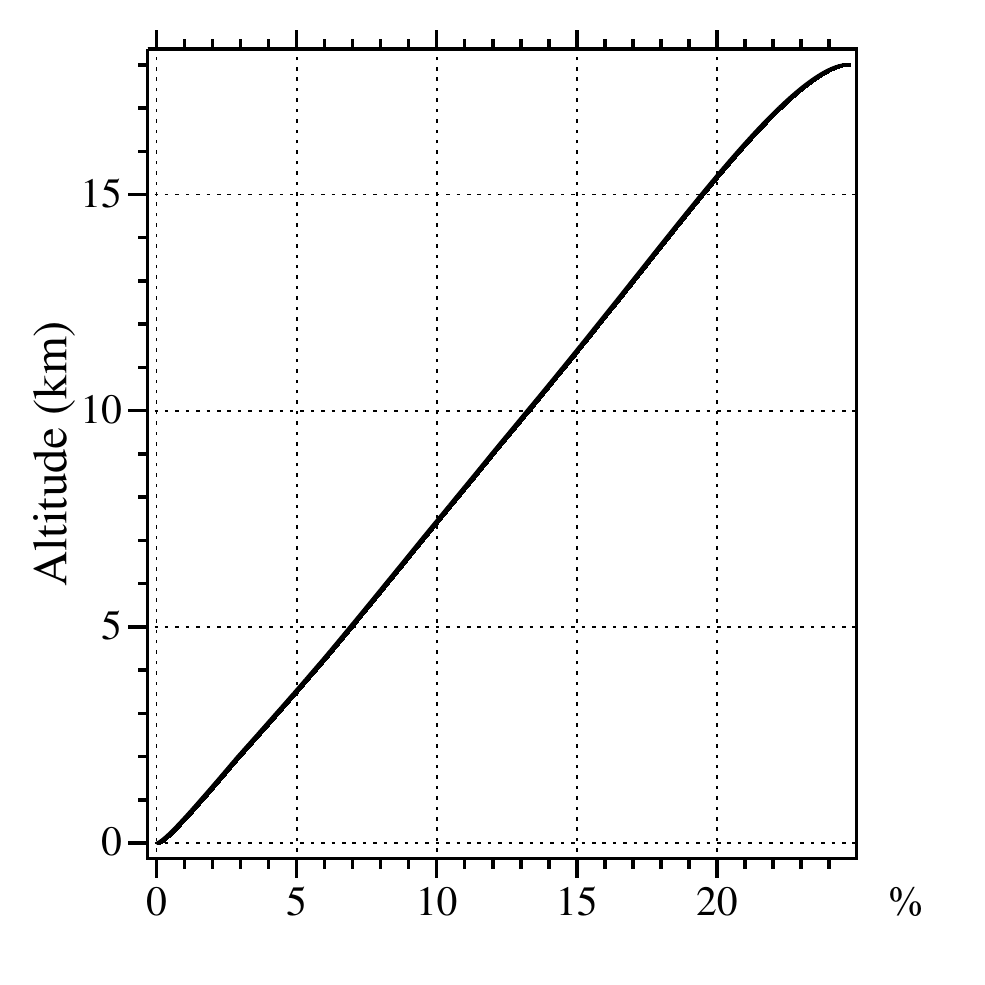}
\caption{Condensed fraction of steam in the baseline simulation.}
\label{fig:Condens}
\end{figure}

Figure \ref{fig:Condens} shows the condensed fraction of the
steam. The condensed fraction increases rather steadily all the time
as a function of altitude.

\begin{figure}[htpb]
\centering
\includegraphics[width=3in]{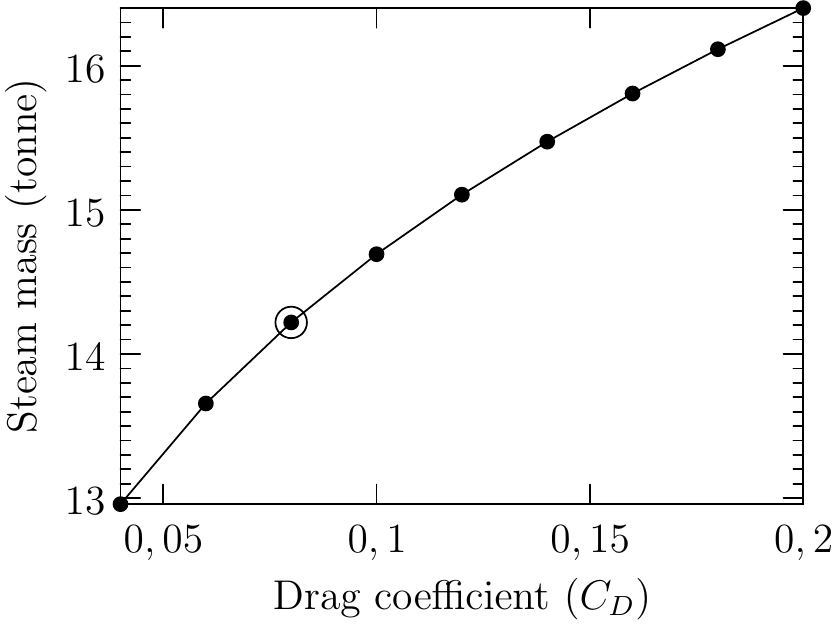}
\caption{Dependence of the required steam mass on the assumed drag coefficient.\goodbreak
The open circle refers to the baseline simulation.}
\label{fig:CD}
\end{figure}

Next, we study the sensitivity of the required initial steam mass to
the various parameters and assumptions. The effect of the aerodynamic
drag coefficient on the required steam mass is shown in
Fig.~\ref{fig:CD}. If the balloon would be a sphere ($C_D \approx 0.2$
at relevant high Reynolds number) instead of drop-shaped, 2.2 tonnes
more steam would be required. The drag coefficient is rather important
because the smaller it is, the faster the balloon goes up, which
shortens the time available for skin cooling. The effect is somewhat reduced
by the fact that a higher velocity increases the convective cooling
rate, however. \MARKI{In any case, the
  overall feasibility of the concept does not hinge on aerodynamics
  since even if the balloon would be cube-shaped
  with $C_D=1.0$, we calculated that the steam
mass would still be only 23 tonnes, i.e., a steam to payload mass ratio of
2.3 instead of the baseline value of 1.4.} 

\begin{figure}[htpb]
\centering
\includegraphics[width=3in]{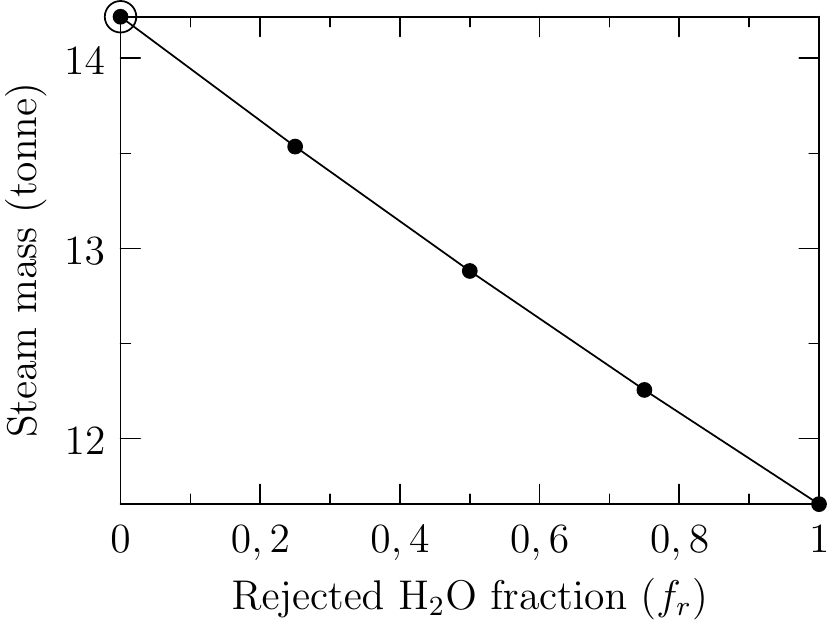}
\caption{Dependence of the required steam mass on the rejected fraction of condensed water.}
\label{fig:condrej}
\end{figure}

If the condensed water is rejected partially or fully, the performance
can be improved further (Fig.~\ref{fig:condrej}). The effect is nearly linear
in the rejected fraction and it is of comparable magnitude to the
difference between the droplet and spherical shape of the balloon.

\begin{figure}[htpb]
\centering
\includegraphics[width=3in]{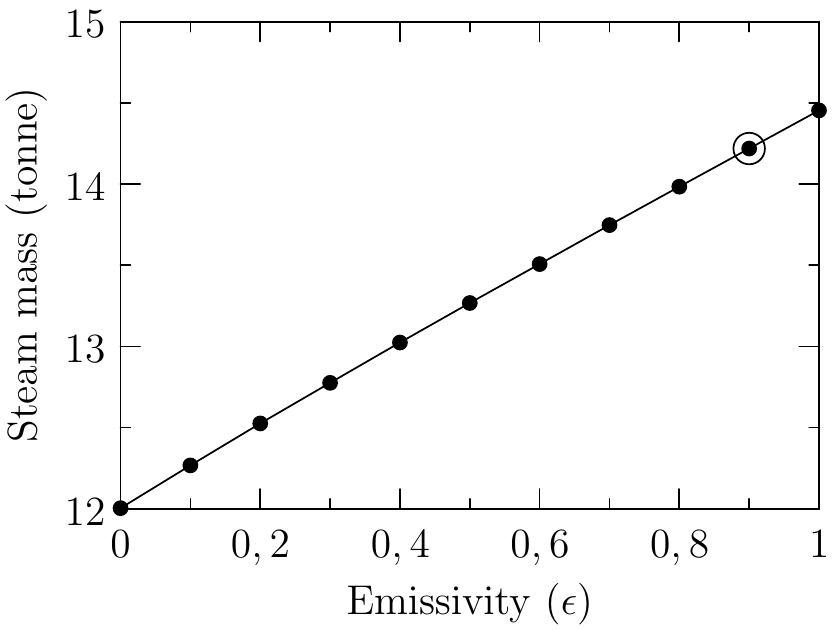}
\caption{Dependence of the required steam mass on the thermal emissivity of the balloon skin.}
\label{fig:emiss}
\end{figure}

If one coats the balloon by shiny metal, its thermal emissivity
becomes nearly zero. The magnitude of the effect is similar to that of
rejecting the condensed water (Fig.~\ref{fig:emiss}). \MARKII{A
  coating and other insulation options are discussed in section
  \ref{sect:insulation} below.}

\begin{figure}[htpb]
\centering
\includegraphics[width=3in]{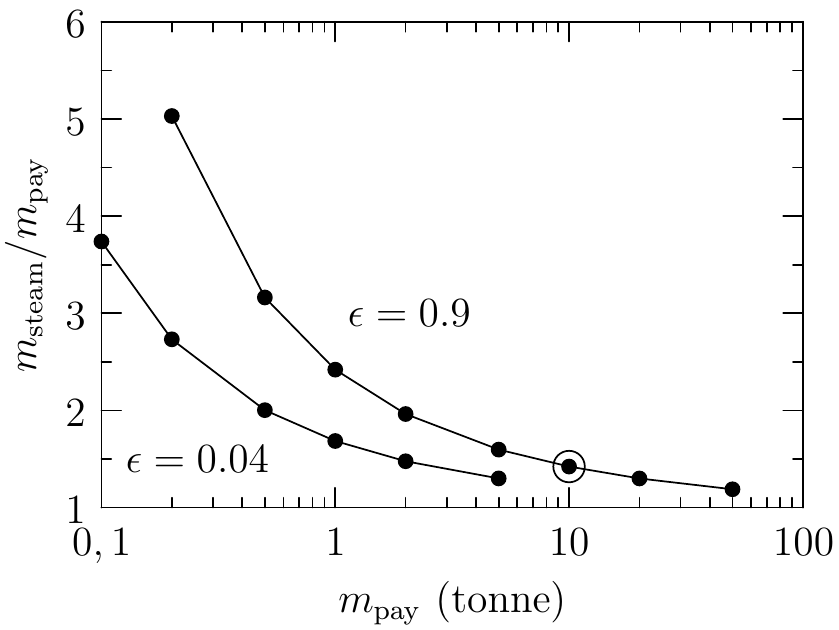}
\caption{Steam per payload mass, as function of the payload.
  The lower curve shows the metal-coated case.}
\label{fig:mass}
\end{figure}

Thus far we have assumed that the payload is 10 tonnes. The effect of
the payload mass on the steam to payload ratio
($m_\mathrm{s}/m_\mathrm{pay}$ ratio) is shown in
Fig.~\ref{fig:mass}. When the payload is smaller, one needs more
steam, relatively speaking, because the skin cooling is
more important in relative terms. The steam to payload ratio is $1.45$ for
a 10 tonne payload, but $5.15$ for a 200 kg payload. Figure \ref{fig:mass}
also shows the simulation with a metal-coated balloon where the
emissivity is assumed to be $\epsilon=0.04$. With metal coating, it
is possible to go down to 100 kg payload while keeping a reasonable
steam to payload ratio of less than 4.

\begin{figure}[htpb]
\centering
\includegraphics[width=3in]{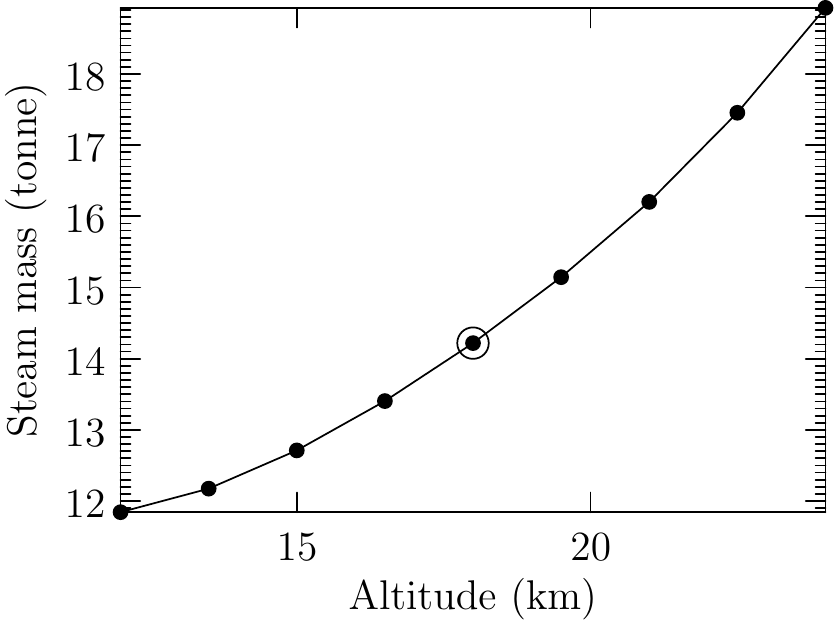}
\caption{Dependence of the required steam mass on the target altitude.}
\label{fig:alt}
\end{figure}

For reference purpose for tradeoff studies with rocket vehicle
design, in Fig.~\ref{fig:alt} we show the effect of the target
altitude on the required steam mass. Qualitatively, for a given
orbital payload mass, a higher launch altitude reduces the rocket's
launch mass, but increases the steam and balloon envelope mass
relative to the rocket's launch mass.

\begin{figure}[htpb]
\centering
\includegraphics[width=3in]{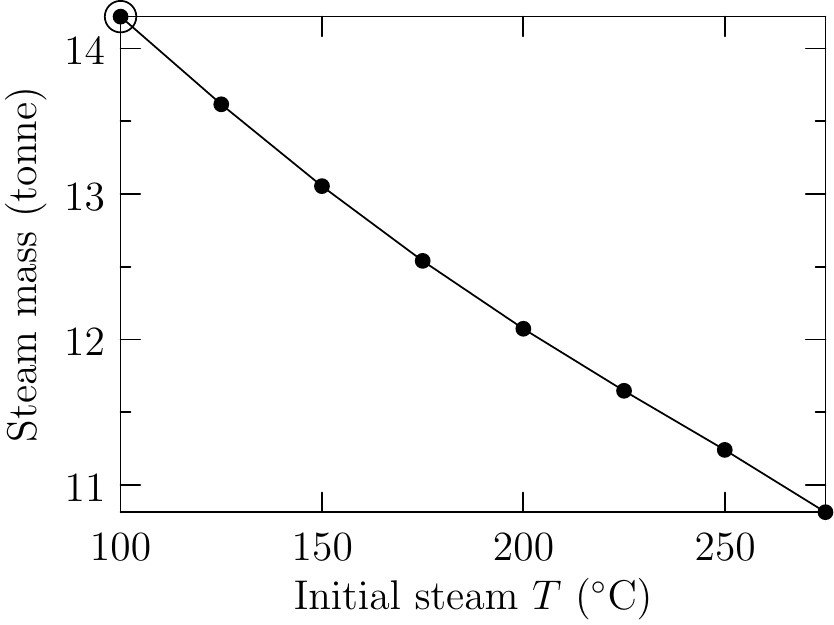}
\caption{Dependence of the required steam mass on the initial steam temperature.}
\label{fig:SteamT}
\end{figure}

Finally, in Fig.~\ref{fig:SteamT} we show the effect of the initial
steam temperature on the required steam mass. The required mass of the
steam gets smaller if the initial temperature is increased above the
boiling point. The effect is not dramatic, but nevertheless
superheating the steam to some extent is a possible approach if the
balloon material tolerates such temperatures. \MARKII{For example, nylon 66 which
  is the material used to make hot air balloon envelopes tolerates a few hours
  of exposure to 180$^{\circ}$\,C, although weakens mechanically}. An additional minor practical
motivation for using some superheating may be to reduce the amount of
condensed hot water that one has to take care of during filling.

\section{THREE DESIGN VARIANTS}
\label{sect:variants}

\begin{figure}[htpb]
\centering
\includegraphics[width=3in]{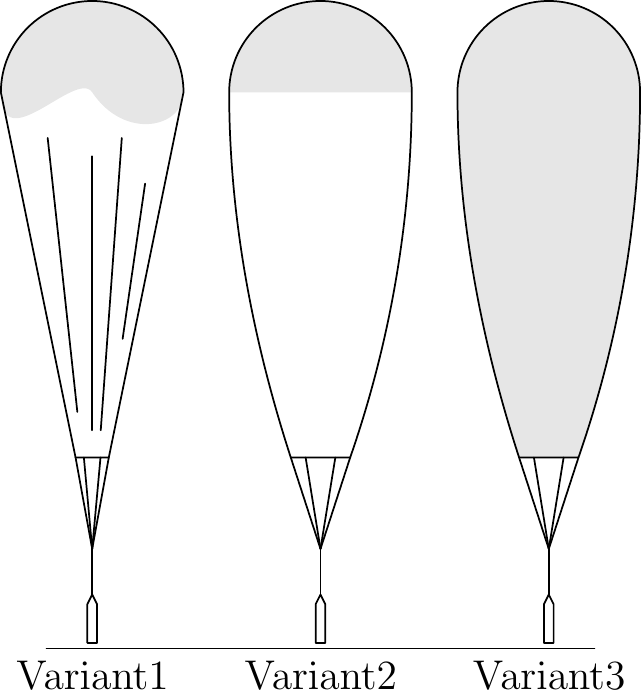}
\caption{The three design variants of the balloon launch-ready on
  ground.\goodbreak  Steam is shown as grey. In variant 2 the lower part of the
  balloon is filled by warm air.}
\label{fig:schematic}
\end{figure}

As the balloon rises, the steam expands. We discuss three design
variants (Fig.~\ref{fig:schematic}), \MARKI{all based on non-elastic
  balloon fabric,} that accommodate the steam expansion in different ways:
\begin{enumerate}
\item Fill the balloon partially with steam. When the balloon
  rises, the steam gradually expands and at the target altitude it fills the
  balloon completely. This corresponds to the so-called zero pressure gas balloon\cite{RainwaterAndSmith2004}.
\item Fill the upper part of the balloon with steam and the lower part
  with warm air. When the balloon rises, the steam expands and
  displaces the air which exits from an opening or openings at the
  bottom.
\item Fill the balloon completely with steam. When the balloon rises,
  excess steam continuously exits from opening(s) at the
  bottom. The exited steam tends to rise faster than the balloon
  itself because it is not slowed down by the payload. The balloon is surrounded by
  a large rising steam cloud from all sides.
\end{enumerate}

Variant 1 is the simplest to implement because it requires no excess
steam and no warm air. A potential drawback is the somewhat
non-aerodynamic shape of the loosely hanging lower part of the
balloon. Freezing of the condensed water in the lower part can
presumably be avoided by geometric design.

Variant 2 also needs no excess steam, but it needs a source of warm
air and filling channels for it. The bulging caused by the warm air
guarantees a good aerodynamic shape, and the warm air also contributes
to the lift in the tropospheric portion of the climb (Fig.~\ref{fig:Speed}). The steam and
air layers have a large density difference which evidently keeps them
well separated. The warm air cools adiabatically. Therefore, in the
troposphere its temperature difference with the ambient stays roughly
constant or somewhat decreases. In the stratosphere, the air inside
the balloon continues to cool adiabatically. If initially at +100
$^{\circ}$C, the internal air cools below the freezing point at 8.3 km
and below the ambient stratospheric temperature of -55 $^{\circ}$C at
13.6 km altitude. Accordingly, the internal air should be pushed out
by the steam before 13.6 km, because above this altitude the air would
contribute a negative lift. Thus, the initial amount of steam should
be selected so that the steam fills the balloon completely at 13.6 km.
Then, between the 13.6 km and 18 km altitudes, the gas that gets
pushed out from the bottom due to adiabatic expansion is no longer
air, but steam.\footnote{This calculation is conservative regarding
  performance, because the internal air was assumed to have zero
  humidity. In reality the absolute humidity of the internal air would
  be the same as that of local air on ground and so it would cool more
  slowly than completely dry air.}

Variant 3 needs several times more steam. Before leaving from the
bottom, the extra steam contributes to the lift directly. After
going out, the exited steam tends to rise faster than the balloon, so it
contributes to the lift by providing an upward tailwind. The exited
steam is also warmer than the surrounding air, thus decreasing the
convective cooling of the skin. The cloud that forms around the
balloon may also decrease the radiative cooling rate. Expectedly,
variant 3 should be able to squeeze out the maximum lifting
performance out of a balloon envelope of a given size. The drawback of
variant 3 is that it needs more steam than the other variants. How the
exited steam interacts with the balloon would be an interesting topic
for experimental research.

Variants 2 and 3 are rather well in line with the aerodynamic
assumptions made in the simulation: the balloon has a constant
streamlined droplet shape. The contribution to the lift by the warm
air of variant 2 and the extra steam of variant 3 were not included in
the simulation, however. The simulation therefore represents an
intermediate case between variants 1 and 2. Simulating variants 1, 2
and 3 in detail is beyond the scope of this paper.

\section{THE EFFECTS OF THERMAL INSULATION AND AIR ADDED IN STEAM}
\label{sect:insulation}

Thus far we have assumed that the balloon has no thermal insulation
except that the effect of reducing the infrared emissivity of the skin
was explored in Fig.~\ref{fig:emiss}. We found (Fig.~\ref{fig:mass})
that surface cooling starts to reduce performance when the size of the
balloon is such that its natural payload is less than a few
tonnes. Thus, for lifting rockets relevant for orbital
launches, surface cooling is typically not important and thermal
insulation need not be considered. If one wants to extend
applicability towards smaller payloads such as atmospheric science
instruments, however, consideration of thermal insulation becomes
worthwhile.

A straightforward way to reduce thermal emissivity would be to add a metal
coating to one or both sides of the envelope. \MARKII{Such coating
  would not add much weight because the aluminium coating that is used,
e.g., on mylar and kapton plastic membranes used in space
technology is typically only 0.1 $\mu$m thick.} However, because the
envelope is large, such coating may be costly. Metal coating is
typically not used in hot-air balloons, for example.

Both the convective and radiative heat loss could be reduced by a double
or multiple wall structure where some gas, possibly together with some
thermal insulation material, exists between the inner and outer
walls. However, using insulation increases the manufacturing cost and
the packaged volume requirement and may make folding of the material
more tricky.

If the balloon contains pure steam, the gas temperature is uniform and
equal to the boiling point of water at the prevailing pressure. If the
steam contains some admixture of air, then it is possible that a layer
of cooler and drier air forms at the walls, because the cool wall
condenses out water from the adjacent layer of gas.  As a result, the
partial pressure of water vapour is lower near the walls although the
total pressure is uniform.  Then the wall temperature is lower and the
surface energy loss is reduced. Thus, the addition of a small fraction
of air in the steam might ideally lead to an automatic formation of an
insulating air layer at the walls.

Somewhat unfortunately for this idea, a cooler and drier air layer
near the wall is denser than the almost pure steam which fills the
bulk of the balloon's volume. Therefore a convective cell forms where
the wall air layer sinks and the steam in the middle of the balloon
rises. Such convection tends to reduce the benefits of the insulating air
layer. The cool walls condense out water from the gas and the produced
air sinks to the bottom. In variants 2 and 3, the air exits from the
bottom because it gets pushed out by the adiabatic expansion of the
balloon gases.

If the wall temperature drops below the freezing point in this scheme,
ice starts to form at the walls, although in the bulk of the balloon
temperature remains well above freezing and nucleation droplets within
the steam are in liquid state. Water immobilised as ice contributes to
the dead mass, but the associated performance penalty is only
moderate, because the corresponding penalty was not too significant even if all the
condensed water would remain in the system (Fig.~\ref{fig:condrej}).

If the wall temperature is successfully reduced by an insulating air
layer, the heat loss of the skin is reduced.

The addition of air in the steam reduces its lifting capability, so it
is not self-evident that the method can produce a net benefit.
However, if an air fraction range exists where a net benefit is
produced, applying the method is attractive because the technical
implementation is easy.

\section{DISCUSSION AND CONCLUSIONS}

Using steam to fill a balloon seems a promising concept for lifting
rockets into the stratosphere for launch. Steam is cheaper than helium or
hydrogen, and it is safe. Unlike helium, steam is a long-term
sustainable and scalable option. Unlike a hot-air balloon, the
ground-filled steam balloon carries no hardware but the payload.

Our baseline mission scenario is that the balloon is launched so that
the wind moves it over the sea. The balloon rises swiftly to the launch
altitude where the rocket is separated and launched. Then the steam is
released by making an opening at the top of the balloon, for example
by a pyrotechnic fuse wire that was weaved onto the balloon
envelope. Alternatively one can keep the balloon intact so that after
releasing the payload, it climbs above $\sim 30$ km where the steam
freezes onto the envelope forming a $\sim 0.6$ mm thin layer of
ice. Then the balloon falls into the sea, from where it can be
recovered. The balloon may be reused for another flight or
the materials can be recycled. Because the lifting
phase is relatively fast (typically of order ten minutes), potential damages like
a small hole in the envelope are not harmful and do not prevent
reuse. The condensed water that trickles down the envelope is allowed
to drop down. The water should be managed so that it does not drop onto the
payload.


The main simplifications and approximations made in the
present simulation were the following:
\begin{enumerate}
\item For calculating the aerodynamic drag, a constant shape of the
  balloon was assumed
  (characterised by the equivalent spherical radius $r$ of the initial steam
  volume and the drag coefficient $C_D$). In reality, 
  the steam expands as the balloon rises.
\item We estimated the cooling surface area from the steam volume $V$
  and the balloon radius $r$ by assuming that the balloon is a
  vertical cylinder. While the balloon's shape is not cylindrical in reality,
  the approximation correctly represents the fact that as the steam volume
  expands, the cooling surface area increases.
\end{enumerate}

For lifting a few tonne of larger payloads into the stratosphere,
the effect of surface cooling is rather unimportant. If one
wants to extend the applicability to smaller payloads such as
atmospheric science instruments, thermal insulation may become
relevant. We speculatively discussed the possibility that mixing
some air in the steam might improve insulation by creating a
cooler air layer adjacent to the walls.

In summary, a steam balloon that is filled on ground looks an
attractive option for lifting tonne scale and larger payloads such as
rockets into the stratosphere. For payloads of such size, a standard
plastic balloon skin without any thermal insulation works well.  Steam
is inexpensive, safe and environmentally friendly, and the steam
balloon is comparable to a helium balloon in simplicity and
operational safety. Experimental verification of the concept could be
performed easily by utilising a standard hot-air balloon envelope and
a truck-mounted portable steam generator for the filling operation,
\MARKII{and releasing a payload would not be mandatory}. \MARKII{On
  the modelling side, numerical 3-D aerodynamic simulations could
  be performed, including the modelling of different ways of suspending and deploying
  the payload.}

\section*{ACKNOWLEDGEMENTS}
The authors thank Perttu Yli-Opas
and Johan K\"utt for fruitful discussions. The results presented
have been achieved under the framework of the Finnish Centre of
Excellence in Research of Sustainable Space (Academy of Finland grant
number 312356).



\section*{APPENDIX}

\section{ESTIMATING THE CONVECTIVE COOLING RATE}

We estimate the convective cooling rate at the outer surface of the balloon by using a
correlation function of heat transfer for forced convection around a
sphere \cite{Whitaker1972}. The convective heat transfer is
\begin{equation}
u_\mathrm{conv} = \frac{k \Delta T}{L} \mathrm{Nu}_L
\end{equation}
where $k$ is the thermal conductivity of air, $\Delta T$ the
temperature difference between the steam and the atmosphere, $L=2r$ the
length scale and $\mathrm{Nu}_L$ the Nusselt number corresponding
to the length scale $L$ and given by
\begin{equation}
\mathrm{Nu}_L = 2 +
\left(
0.4\,\mathrm{Re}_L^{1/2} + 0.06\,\mathrm{Re}_L^{2/3}
\right)
\mathrm{Pr}^{0.4} \left(
\frac{\mu}{\mu_s}
\right)^{1/4}.
\label{eq:Whitaker}
\end{equation}
Here $\mathrm{Re}_L = L \varv/\nu$ is the Reynolds number, $\nu$ the
kinematic viscosity, $\mathrm{Pr}=\nu/\alpha$
the Prandtl number, $\alpha$ the thermal diffusivity and
$\mu=\rho\nu$ the dynamic viscosity of ambient air, while $\mu_s$
is the dynamic viscosity of air at the surface temperature of the
balloon.

There is some risk in using Eq.~(\ref{eq:Whitaker}) because it was
validated only up to $\mathrm{Re}_L=7.6\cdot 10^4$\cite{Whitaker1972}, while in our case
$\mathrm{Re}_L\sim 10^7$. Also, our balloon is not spherical, but has a
more streamlined shape. However, we believe that
Eq.~(\ref{eq:Whitaker}) gives the correct order of magnitude.


\end{document}